\documentclass[prl,superscriptaddress,citeautoscript,a4paper,reprint]{revtex4-1}
\pdfoutput=1
\usepackage{amsmath,amsfonts,amssymb}
\usepackage{newtxtext,newtxmath}
\usepackage[pdftex]{graphicx}
\usepackage{microtype}
\usepackage{bm}

\usepackage[svgnames]{xcolor}
\usepackage[
	colorlinks=True,linkcolor=DarkRed,citecolor=ForestGreen,urlcolor=MediumBlue,
	pdfstartview=FitH,bookmarks=False,pdfpagemode=UseNone
]{hyperref}

\begin{document}

\title{Modulation of superconductivity by quantum confinement in doped strontium titanate}

\author{D. Valentinis}
\affiliation{Department of Quantum Matter Physics (DQMP), University of Geneva, 24 quai Ernest-Ansermet, 1211 Geneva 4, Switzerland}
\author{Z. Wu}
\email[Corresponding author. E-mail: ]{zhenpingwu@bupt.edu.cn}
\affiliation{State Key Laboratory of Information Photonics and Optical Communications \& School of Science, Beijing University of Posts and Telecommunications, Beijing 100876, China}
\author{S. Gariglio}
\affiliation{Department of Quantum Matter Physics (DQMP), University of Geneva, 24 quai Ernest-Ansermet, 1211 Geneva 4, Switzerland}
\author{D. Li}
\affiliation{Stanford Institute for Materials and Energy Sciences, SLAC National Accelerator Laboratory, Menlo Park, California 94025, USA}
\author{G. Scheerer}
\affiliation{Department of Quantum Matter Physics (DQMP), University of Geneva, 24 quai Ernest-Ansermet, 1211 Geneva 4, Switzerland}
\author{M. Boselli}
\affiliation{Department of Quantum Matter Physics (DQMP), University of Geneva, 24 quai Ernest-Ansermet, 1211 Geneva 4, Switzerland}
\author{J.-M. Triscone}
\affiliation{Department of Quantum Matter Physics (DQMP), University of Geneva, 24 quai Ernest-Ansermet, 1211 Geneva 4, Switzerland}
\author{D. van der Marel}
\affiliation{Department of Quantum Matter Physics (DQMP), University of Geneva, 24 quai Ernest-Ansermet, 1211 Geneva 4, Switzerland}
\author{C. Berthod}
\affiliation{Department of Quantum Matter Physics (DQMP), University of Geneva, 24 quai Ernest-Ansermet, 1211 Geneva 4, Switzerland}

\begin{abstract}

Quantum confinement in a thin-film geometry offers viable routes for tuning the critical properties of superconductors through modification of both density of states and pairing interaction. Low-density systems like doped strontium titanate are especially susceptible to these confinement-induced effects. In this paper, we show that the superconducting critical temperature $T_c$ is enhanced through quantum confinement in SrTiO$_3$/SrTi$_{1-x}$Nb$_x$O$_3$/SrTiO$_3$ heterostructures at $x=1\%$ concentration, by measuring resistivity transitions and the Hall carrier density for different thicknesses of the doped layer. We observe a nonmonotonic raise of $T_c$ with decreasing layer thickness at constant carrier density as estimated from the Hall effect. We analyze the results by solving a two-band model with a pairing interaction reproducing the density-dependent $T_c$ of doped SrTiO$_3$ in the bulk, that we confine to a potential well established self-consistently by the charged Nb dopants. The evolution of the theoretical $T_c$ with thickness agrees well with experiments. We point out the possible role of density inhomogeneities and suggest novel methods for engineering superconductivity in epitaxial thin films. 

\end{abstract}

\maketitle

\section{Introduction}

Quantum confinement has emerged as an effective tool to tune the critical properties of superconductors by reducing the system size along any spatial coordinate down to nanometric scale, thereby tailoring the electronic structure through quantization. Specifically, confining a three-dimensional bulk along one coordinate generates a quasi-two dimensional (2D) band structure composed of a series of 2D subbands. This alters the critical temperature $T_c$ through the modification of both the density of states (DOS) and the pairing interaction, which can lead to substantial $T_c$ enhancements with respect to the bulk. Motivated by this prospect and supported by refinements in the thin-film growth techniques, several groups have recently obtained high-quality experimental data on epitaxial superconducting thin films, which shed new light on confinement-induced $T_c$ modifications: Nb films grown on SiO$_2$ and Al$_2$O$_3$ substrates showed a decrease of $T_c$ in the thin-film limit, attributed to inverse proximity effect from the substrate \cite{Pinto-2018}, while spectacular enhancements up to more than four times the bulk value were found in TaS$_2$ flakes \cite{Yang-2018}, due to the suppression of a competing charge-density wave phase. Doped strontium titanate SrTiO$_3$ (STO) offer a unique opportunity to study confinement effects on superconductivity in the absence of substrate interactions or competing phases. In fact, STO can be grown in the form of thin film and shows a superconducting ground state down to very low carrier density \cite{Schooley-1965, Koonce-1967, Lin-2013, Lin-2014}, the latter being finely controlled by chemical doping. Besides, its low carrier density favors confinement-induced $T_c$ enhancements due to shape resonances \cite{Valentinis-2016-2}. This interplays with the multiband and possibly unconventional nature of superconductivity in STO \cite{vanderMarel-2011, Lin-2014}, which is intimately connected to peculiar Fermi-surface properties \cite{Lin-2013, Lin-2014}. 

Stoichiometric STO is a band insulator with perovskite crystal structure and cubic symmetry at room temperature, which undergoes a tetragonal distortion below 105~K. The material becomes metallic at carrier densities as low as $5.5\times10^{17}~\mathrm{cm}^{-3}$ upon chemical doping \cite{Lin-2013, Lin-2015a}. The doped carriers populate up to three conduction bands of mainly Ti $3d$ character \cite{vanderMarel-2011, Lin-2014}. The nature of the metallic state has raised questions, as it displays Fermi-liquid phenomenology down to the lowest doping \cite{vanderMarel-2011, Lin-2013} despite the ineffectiveness of conventional momentum-relaxation mechanisms \cite{Lin-2015a, McCalla-2018}. Strong electron-phonon coupling manifests itself in polaronic effects \cite{vanderMarel-2011, Devreese-2010, vanMechelen-2008, Chen-2015, Wang-2016, Verdi-2017}. STO is also a quantum paraelectric in which the proximity to a ferroelectric instability leads to an unusual enhancement of dielectric screening at low temperature.

Upon cooling below $T \approx 300$~mK, the system enters a superconducting state with a dome-shaped critical temperature as a function of carrier concentration \cite{Schooley-1965, Koonce-1967, Lin-2014}. The mechanism of this superconductivity, which occurs in the anti-adiabatic regime where the Fermi temperature is less than the Debye temperature \cite{vanderMarel-2011, Lin-2013, Lin-2014, Gorkov-2016, Valentinis-2017} and exhibits an inverted isotope effect \cite{Stucky-2016}, is a subject of ongoing debate. Phonon-mediated pairing has been proposed, invoking a soft mode related to the antiferrodistortive transition taking place at 105~K \cite{Appel-1969}, high-frequency longitudinal optical phonons \cite{Klimin-2016, Gorkov-2016}, or an optical mode associated with the ferroelectric quantum critical point \cite{Edge-2015, Wolfle-2018}. Plasmonic mechanisms were also considered \cite{Takada-1980, Ruhman-2016}. Recently, $T_c$ enhancements in the density range $n \approx 10^{18}$--$10^{19}$ cm$^{-3}$ were obtained through Ca substitution in oxygen-reduced samples, revealing the coexistence and interplay of ferroelectricity and superconductivity \cite{Rischau-2017}. The roles of different bands is also debated: early tunneling measurements showed two gaps \cite{Binnig-1980}, one of which recently ascribed to surface superconductivity \cite{Eagles-2018}. Thermal conductivity measurements found multiple nodeless gaps \cite{Lin-2014b}, while microwave spectroscopy results are compatible with single-gap electrodynamics \cite{Thiemann-2018}.

Evidence for confinement effects on $T_c$ in STO-based heterostructures first emerged in transport experiments on LaAlO$_3$/SrTiO$_3$, which hosts a quasi-2D superconducting electron liquid extending on the STO side of the interface \cite{Reyren-2007, Caviglia-2008, Gariglio-2015b}. There, electrostatic field-effect modulation of the confinement potential reveals a critical temperature drawing a dome as a function of gate voltage with maximum $T_c$ comparable with the bulk $T_c$ \cite{Gariglio-2016, Valentinis-2017}. Remarkably, superconducting systems composed solely of STO are also affected by quantum confinement \cite{Kim-2012}. Quasi-2D heterostructures consisting of a Nb-doped SrTi$_{1-x}$Nb$_x$O$_3$ (Nb:STO) slab at $x=1$\% grown between pristine STO layers show a $T_c$ increase by up to a factor 1.5 between the thickest and thinnest slabs.

Motivated by the work of Kim \textit{et al}.\ \cite{Kim-2012}, we have investigated the possibility that such $T_c$ increase may be due to shape resonances. Because the $T_c$ of STO depends on carrier density, it is important to precisely monitor the carrier density of the Nb:STO slabs at the various thicknesses, before ascribing the variations of $T_c$ to quantum confinement. We have therefore grown a new set of STO/Nb:STO/STO$_{\mathrm{sub}}$ heterostructures, where STO$_{\mathrm{sub}}$ denotes the STO substrate, and measured both their $T_c$ and Hall carrier density as a function of the Nb:STO layer thickness.

\section{Transport experiments on $\text{Nb}$:STO thin films}

\begin{figure}[tb]
\includegraphics[width=\columnwidth]{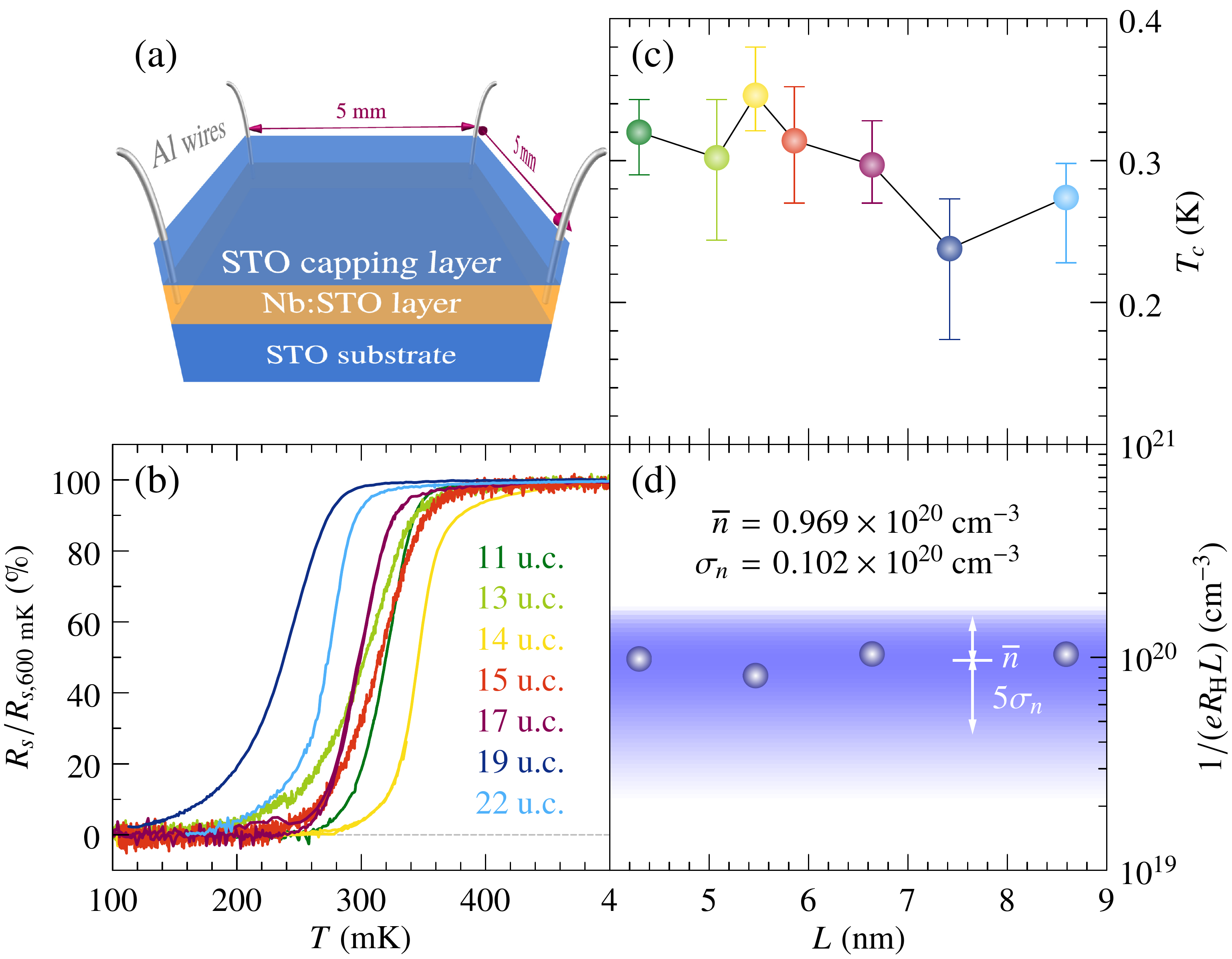}
\caption{\label{fig:exp}
Transport measurements on Nb:STO thin films. (a) Sketch of the device and geometry of the transport measurements. (b) Sheet resistance $R_s$ normalized to the value at 600~mK versus temperature for Nb:STO films of various unit-cell (u.c.) thicknesses. (c) Critical temperature and (d) carrier density of the films at 600~mK as a function of thickness. The thickness $L$ is evaluated as the number of unit cells times the lattice parameter $3.905$~\AA{} of STO. The lower bars, symbols, and upper bars in (c) show the temperature at 10\%, 50\%, and 90\% of the resistive transition, respectively. The background color shading in (d) represents a Gaussian distribution of densities with average $\overline{n}$ and standard deviation $\sigma_n$ as indicated.
}
\end{figure}

A series of Nb:STO films with thicknesses ranging from 1 to 22 monolayers was grown on STO substrates using pulsed laser deposition. Sample fabrication and measurement details can be found in the Methods section. Four-point resistance measurements were performed in a standard dilution cryostat [see Fig.~\ref{fig:exp}(a)]. Nb:STO layers of 8 unit cells (u.c.) and less all showed insulating behavior with over $10~\mathrm{M\Omega}$ resistance. For 11~u.c.\ and more, a single superconducting transition was observed when the samples were cooled below a thickness-dependent critical temperature, as shown in Fig.~\ref{fig:exp}(b). In this paper, we define the critical temperature at the 50\% drop of the resistance below the normal-state value at 600~mK. Figure~\ref{fig:exp}(c) shows $T_c$ for all superconducting samples as a function of the Nb-doped layer thickness. Top and bottom error bars correspond to the temperatures at 90\% and 10\% of the resistive transition, respectively. The 10\%--90\% transition widths are in the range 50--100~mK for all samples. We observe a $T_c$ maximum at $L=5.5$~nm, consistently with the previous report \cite{Kim-2012}. Three-dimensional (3D) carrier densities for different samples were then measured to identify whether the observed dependence of $T_c$ on thickness is solely due to quantum confinement. Figure~\ref{fig:exp}(d) shows the carrier density measured at 600~mK via the Hall effect as $n=1/(eR_{\mathrm{H}}L)$ for selected Nb:STO films, where $R_{\mathrm{H}}$ is the measured Hall coefficient and $e$ is the electron charge (see Methods section). Remarkably, $n$ is nearly constant throughout the spanned thickness range with an average of $0.97\times10^{20}~\mathrm{cm}^{-3}$, slightly lower than the nominal 1\% doping corresponding to $1.7\times10^{20}~\mathrm{cm}^{-3}$ \cite{Leitner-1998, Takahashi-2004}. The standard deviation of the measured densities is about 10\% of the average value. In view of the near constancy of $n$, we propose that the observed $T_c$ variations are controlled by the thickness, in agreement with the quantum confinement scenario. To substantiate this claim, we now model the thickness-, confinement potential-, and density-dependent critical temperature $T_c(L,U,n)$ of the superconducting films.

\section{\boldmath Mean-field model for $T_c$ in quasi-2D geometry}

Using a generalization of the BCS theory, we interpret the variations of $T_c$ in our films as a consequence of quantum confinement. The first step is to construct a model that reproduces the density dependence of $T_c$ in bulk STO. We want a model focussing on superconductivity and we let aside all other aspects in the rich physics of doped STO \cite{Collignon-2017, Gariglio-2015a, Collignon-2018, Pai-2018}. 
Given the uncertainties surrounding the superconductivity of STO, we adopt a minimal two-band model with a local density-dependent pairing interaction of unspecified origin. This model has already served for a comparison of the superconducting domes in bulk STO and at the LaAlO$_3$/SrTiO$_3$ interface \cite{Valentinis-2017}. Two parabolic bands mimic the low-energy sector of the STO conduction band \cite{vanderMarel-2011}: the lowest (light) band has un-renormalized mass $0.5m$ while the second (heavy) band has mass $2m$ and lies $2.4$~meV above the former, $m$ being the free-electron mass. The masses are further renormalized by a factor two due to electron-phonon coupling \cite{McCalla-2018}. The pairing is generated with a BCS-like density-dependent interaction $V(n)$ cut off at $\hbar\omega_{\mathrm{D}}=44$~meV \cite{Ahrens-2007} and without inter-band interaction. We solve the gap equation self-consistently for the chemical potential and $T_c$. The value of $V(n)$ is adjusted such as to reproduce the dome-shaped relation $T_c(n)$ of bulk STO.

\begin{figure}[tb]
\includegraphics[width=\columnwidth]{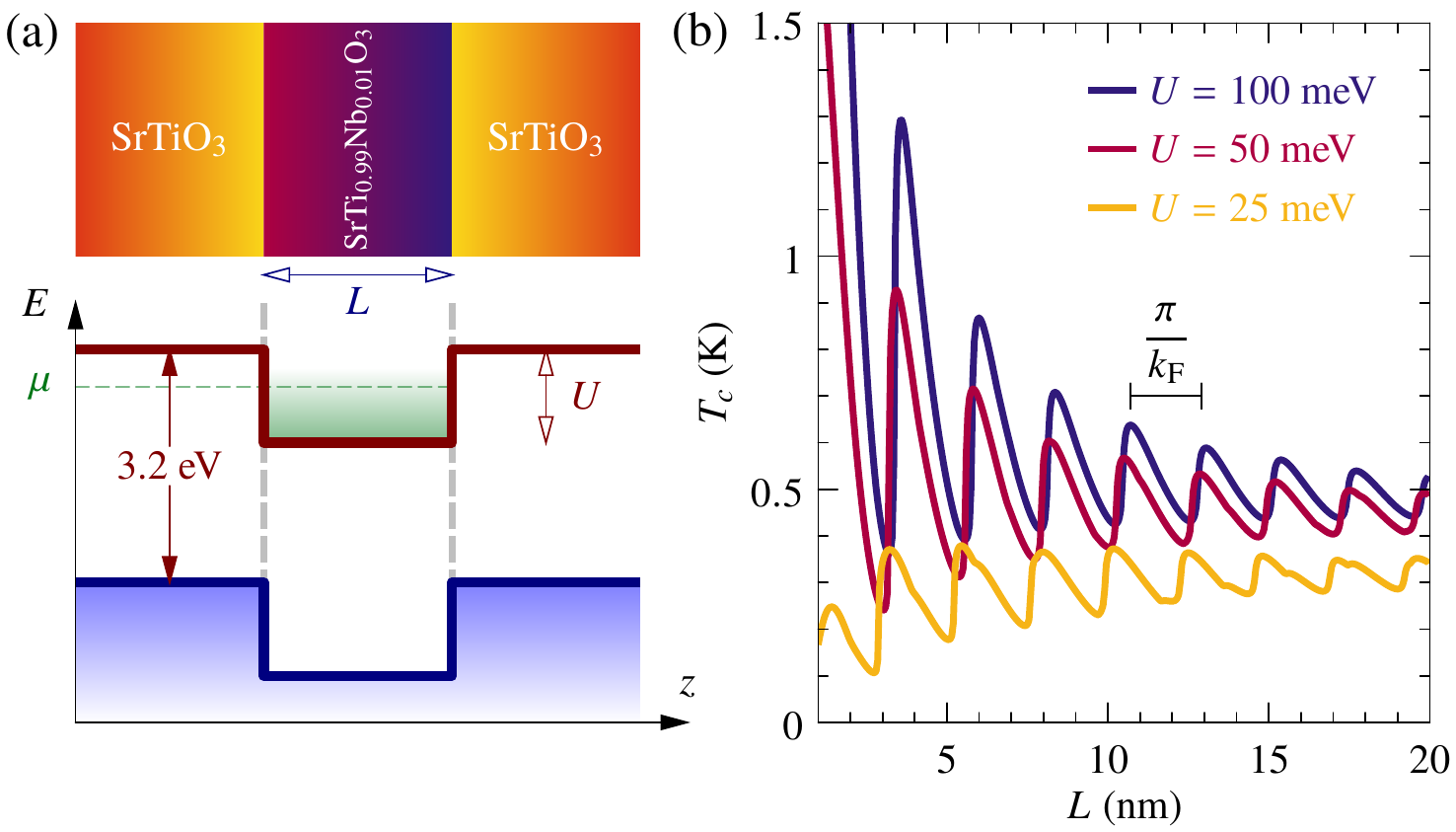}
\caption{\label{fig:model}
(a) Sketch of the STO/Nb:STO/STO heterostructure at $1\%$ Nb and corresponding band alignment diagram. The brown and blue lines mark the conduction and valence bands as a function of growth direction $z$, separated by the bandgap of pristine STO. Doped electrons with chemical potential $\mu$ fill the potential well of depth $U$ (not to scale) created in the doped layer by the electrostatic potential of Nb atoms. (b) Calculated $T_c$ as a function of layer thickness for a density $n=0.97\times 10^{20}~\mathrm{cm}^{-3}$ and various values of $U$. The interval between oscillations equals $\pi/k_{\mathrm{F}}$ with $k_{\mathrm{F}}^3=3\pi^2n$.
}
\end{figure}

The second step is to confine the model in a quasi-2D quantum well of width $L$ and depth $U$ [Fig.~\ref{fig:model}(a)]. The confinement affects the model in three ways. The two 3D bands become series of 2D subbands and, accordingly, the continuous 3D DOS becomes staircase-like. Second, the pairing $V(n)$ becomes a function of $L$ and $U$, and it is different in each subband. Third, all subbands of a given series are coupled owing to the 3D nature of the underlying pairing. These changes modulate $T_c$ with oscillations often termed shape resonances \cite[\& references therein]{Thompson-1963, Blatt-1963, Valentinis-2016-2, Bianconi-2014, Innocenti-2010}. Figure~\ref{fig:model}(b) shows these oscillations of $T_c(L,U,n)$ as a function of $L$ for three values of the confinement potential and a density $n=0.97\times10^{20}$~cm$^{-3}$. At intermediate to strong coupling, the period of the shape resonances is set by the Fermi energy or the interaction cutoff in the adiabatic or anti-adiabatic limit, respectively, according to $\Delta L=\pi\hbar/(\sqrt{2m_b}\sqrt{E_{\mathrm{F}}+\hbar\omega_{\mathrm{D}}})$ for a band of mass $m_b$ \cite{Valentinis-2016-2}. At weak coupling, however, the signatures of the cutoff disappear and the periodicity becomes again $\pi/k_{\mathrm{F}}$, even in the anti-adiabatic limit \footnote{D. Valentinis and C. Berthod, unpublished.}. Our model for STO resides in the weak-coupling anti-adiabatic regime and displays oscillations at $\pi/k_{\mathrm{F}}$. The value of $T_c$ is dominated by the heavy band due to its larger DOS \cite{Valentinis-2017}. It is seen that $T_c$ decreases with decreasing thickness for $U=25$~meV while it increases for 50 and 100~meV. According to Ref.~\onlinecite{Valentinis-2016-2}, the transition between these two behaviors takes place at a critical confinement potential $U^*\approx 2.85\sqrt{\hbar \omega_{\mathrm{D}}Vn}$, which gives 37~meV in our case. $U$ is the sole model parameter that is unknown. The increasing $T_c$ with reducing film thickness in Fig.~\ref{fig:exp}(c) suggests that $U$ should be larger than $U^*$. Further insight may be gained by noticing that the confinement is a result of Nb doping. The substitution of Nb for Ti produces an attractive potential for the electrons, such that the spatial average of this electrostatic potential should correspond to $U$ [see Fig.~\ref{fig:model}(a)]. In the simplest model of a Thomas--Fermi screened Coulomb potential, the spatial average is $(x/a^3)e^2/(\epsilon_0k_{\mathrm{TF}}^2)=x/[N(0)a^3]$, where $x$ is the Nb concentration, $a$ the lattice parameter, $k_{\mathrm{TF}}^{-1}$ the screening length, and $N(0)$ the Fermi-level DOS (without electron-phonon renormalization). Using the zero-temperature value of $N(0)$ in our model, we obtain $U=41$~meV. The best fit discussed below gives a value $U=38.4$~meV fully consistent with this estimate. Note that $U<\hbar\omega_{\mathrm{D}}$, such that a range of extended states outside the well gives a contribution to the pairing \cite{Valentinis-2016-2}.

\begin{figure}[b]
\includegraphics[width=\columnwidth]{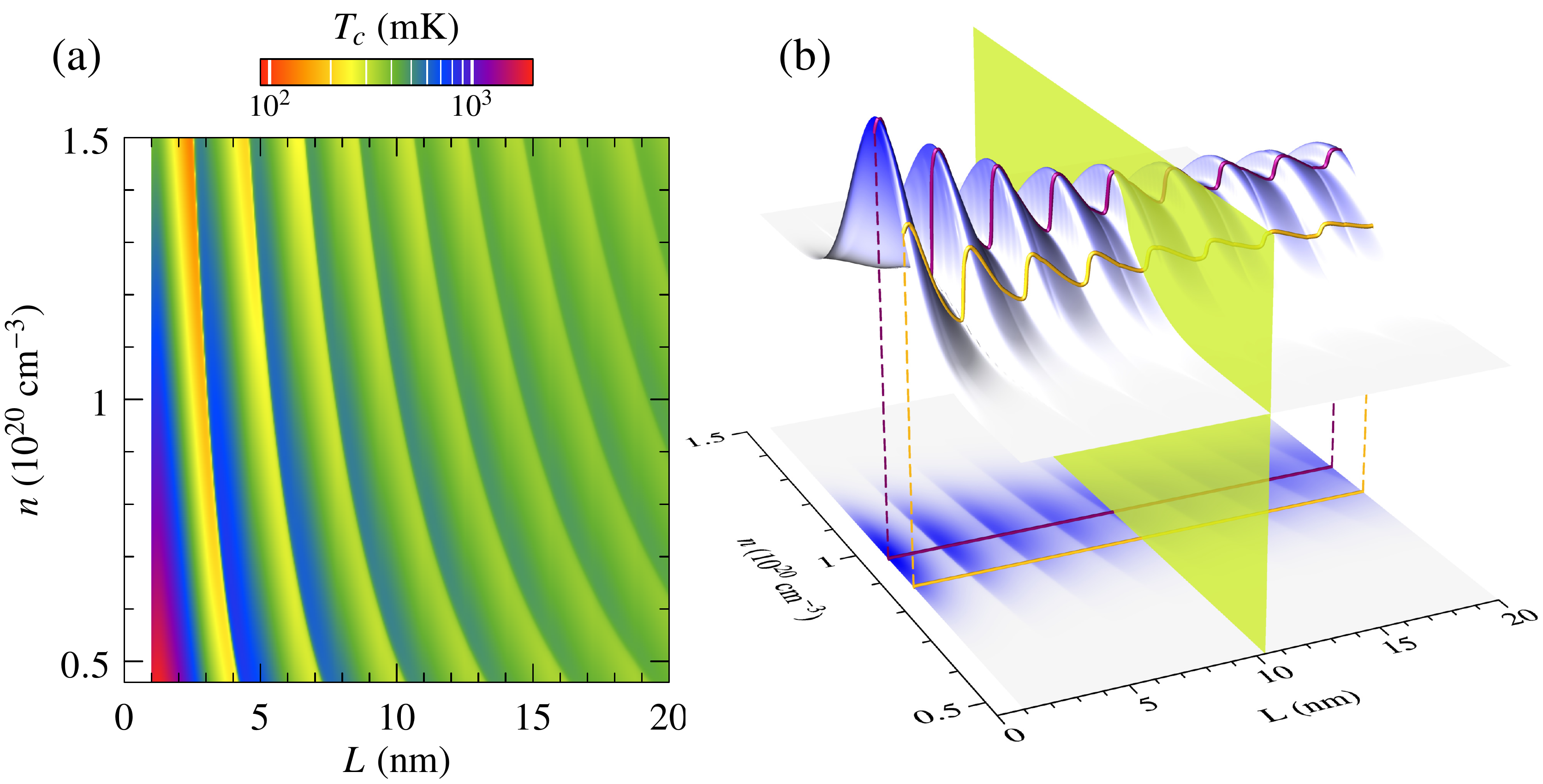}
\caption{\label{fig:map}
(a) Color plot of the calculated critical temperature for a confinement depth $U=38.4$~meV as a function of film thickness $L$ and density $n$. (b) Illustration of Eq.~(\ref{eq:Tc_av}). $T_c$ of panel (a) is multiplied by a normal distribution of width $\sigma_n$ centered at $\overline{n}$. The density average represented by the green plane picks up $T_c$ values at minima (yellow curve) and maxima (purple curve) of shape resonances, thereby smoothing the oscillations.
}
\end{figure}

\begin{figure}[tb]
\includegraphics[width=\columnwidth]{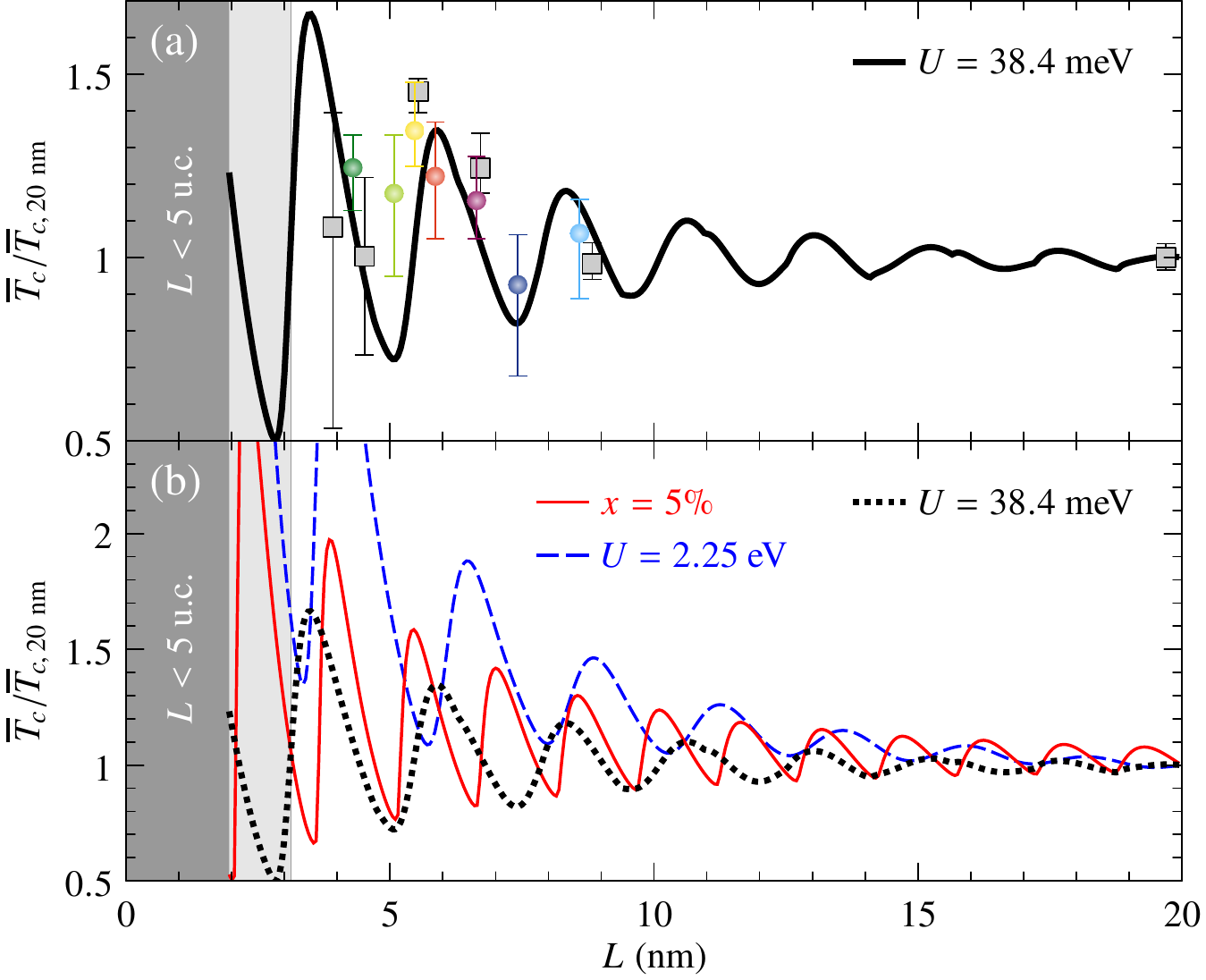}
\caption{\label{fig:fit}
Comparison of the critical temperature measured in Nb:STO films with the predictions of Eq.~(\ref{eq:Tc_av}). (a) Colored circles and gray squares with error bars show the experimental results of this work and data extracted from Fig.~1b of Ref.~\onlinecite{Kim-2012}, respectively, normalized to the $T_c$ value at $L=20$~nm. The width of the symbols corresponds to one lattice spacing. The films are insulating for thicknesses inside the shaded area. The black line is the calculated $\overline{T}\kern-0.15em_c$ for $U=38.4$~meV. We do not expect the model to be reliable for $L\to0$ and report calculations for thicknesses above 5 u.c. (b) Expected behaviors at $x=5\%$ (red) and $U=2.25$~eV (blue-dashed), respectively (see text). The calculated $\overline{T}\kern-0.15em_c$ of panel (a) is shown as a dotted line for comparison.
}
\end{figure}

The simulations in Fig.~\ref{fig:model}(b) overestimate the value of $T_c$ as well as the amplitude of its oscillations compared with the data in Fig.~\ref{fig:exp}(c). The model is designed to approach the $T_c$ of bulk single crystals at large $L$, while the thickest films in Ref.~\onlinecite{Kim-2012} and in the present study consistently show a value lower than the bulk value. In addition to 2D fluctuations and a possible inverse proximity effect from the undoped STO layers, several factors may have an effect on $T_c$ and its oscillations in thin films. A possible phase separation at low density \cite{Caprara-2013} or the disorder present in Nb:STO films \cite{Leitner-1998, Gopalan-1999, Lin-2015b}, which may affect $T_c$ when conjugated with a repulsive interband coupling \cite{Trevisan-2018a, Trevisan-2018b}, are ignored in the model. The scattering time $\tau\sim1$--$2$~ps extracted from our Hall data (see Methods section) is similar to the typical values measured in the bulk \cite{Lin-2013, vanderMarel-2011,Thiemann-2018}, suggesting that disorder effects are not substantially enhanced on going from 3D to quasi-2D. In the following, we ascribe the reduced amplitude of oscillations to inhomogeneities of the carrier density within the films. Such inhomogeneities stem from a charge profile across the film and could also result from nonuniform doping. We account for inhomogeneities by averaging our calculated $T_c$ over a normal distribution of densities. As our only hint at inhomogeneities is the distribution of densities visible in Fig.~\ref{fig:exp}(d), we use the standard deviation of those values as the width of our distribution. In the parameter range of interest, $T_c$ varies relatively slowly with density [see Fig.~\ref{fig:map}(a)]. The variations become faster as $L$ increases, such that inhomogeneities are expected to smoothen the oscillations more efficiently for thicker films. Figure~\ref{fig:map}(b) gives a perspective view of $T_c$ multiplied by the normal distribution of densities. Our final model for the thickness-dependent critical temperature is obtained by averaging this along the density axis:
	\begin{equation}\label{eq:Tc_av}
		\overline{T}\kern-0.15em_c(L,U,n)=\int_0^{\infty} dn'\,
		\frac{e^{-\frac{1}{2}\left(\frac{n'-n}{\sigma_n}\right)^2}}{\sqrt{2\pi}\sigma_n}T_c(L,U,n').
	\end{equation}
We determine the unknown parameter $U$ by fitting Eq.~(\ref{eq:Tc_av}) to the $T_c$ data from Fig.~\ref{fig:exp}(c) and from Ref.~\onlinecite{Kim-2012}, which yields $U=38.4$~meV. The results for $T_c$, normalized to its value at $L=20$~nm, are displayed in Fig.~\ref{fig:fit}(a). $T_c$ increases as $L$ decreases with oscillations of reduced amplitude with respect to Fig.~\ref{fig:model}(b), in semi-quantitative agreement with experiments. Despite of the given 10\%--90\% transition widths, Fig.~\ref{fig:fit}(a) strongly suggests that oscillations due to quasi-2D quantum confinement govern the observed $T_c$ variations in Nb:STO thin films.

\section{Discussion}

A clearcut demonstration of quantum size effects on the critical temperature of thin films is an uneasy task. Several thickness-dependent mechanisms can affect $T_c$, which are not directly related to quantum confinement. 2D fluctuations \cite{Gabay-1993, Benfatto-2009, Singh-2018} will quench $T_c$ in films thinner than the (usually thickness-dependent) coherence length, while disorder and localization also promote an insulating ground state \cite{Finkelstein-1994, Yanase-2009, Caprara-2013, Trevisan-2018b}. Our insulating films of 8~u.c.\ or less presumably realize this type of phenomenology. Besides, inverse proximity effect from a non-superconducting substrate can suppress $T_c$ smoothly as the thickness decreases \cite{McMillan-1968}. If this effect can be subtracted out, quantum oscillations may appear \cite{Pinto-2018}. As these various phenomena all \emph{decrease} $T_c$ in the thin film limit, an enhancement of $T_c$ in thinner films seen in Fig.~\ref{fig:exp}(c) indicates that they are at least not dominant. The superconducting thickness as measured by the angular dependence of the critical field being equal to the film thickness \cite{Kim-2012} is a confirmation that proximity effects are not important in this system. The optimal conditions for observing shape resonances are low-density, weak coupling, and a deep potential well \cite{Valentinis-2016-2}. While previous studies have focused on high-density systems \cite{Orr-1984, Qin-2009, Kang-2011, Navarro-2016, Pinto-2018, Yang-2018}, Nb:STO is in the low-density regime $E_{\mathrm{F}}<\hbar\omega_{\mathrm{D}}$ and moreover has a low coupling in the range 0.1--0.3. The possibility of using the same material, undoped, as a substrate gives an excellent opportunity to achieve atomically flat and controlled epitaxial interfaces. Furthermore, the sandwich structure with a cap layer realizes a symmetric potential well for the films. This is not required for quantum size effects, but makes their interpretation easier, because the subbands are quadratically spaced in energy, and their modeling straightforward by means of a rectangular well. These advantages come at a price: the confinement potential is not very deep.

As the confinement is self-consistently set by the Nb doping, the only way to make it deeper without changing the substrate and capping materials is to increase the doping concentration. Assuming that $n$ scales with $x$, the potential is expected to scale like $U\sim n^{2/3}$ due to increasing screening with increasing $n$. This is to be compared with the critical potential $U^*\sim\sqrt{Vn}$ below which the confinement reduces $T_c$. Considering that $V$ varies with $n$ \cite{Valentinis-2017}, we find $Vn\sim n^\alpha$ with $\alpha\approx0.73$. Our analysis has shown that $U>U^*$ for $x=1\%$; it follows that $U/U^*\sim n^{0.3}$ can be expected to raise upon doping further beyond $1\%$. While this is favorable for quantum oscillations, the density increase itself is unfavorable and a tradeoff takes place. Our simulations show that the confinement wins and the $T_c$ oscillations are slightly enhanced [see Fig.~\ref{fig:fit}(b)] upon scaling the parameters $n$ and $\sigma_n$ by a factor 5 and $U$ by a factor $5^{2/3}$, as appropriate for a doping $x=5\%$.

Another approach to enhance the amplitude of shape resonances is to change substrate: LaAlO$_3$ may be a candidate of choice. According to electronic-structure calculations \cite{Pentcheva-2008}, most of the bandgap difference at the (100) interface between LaAlO$_3$ and SrTiO$_3$ is accommodated by a large conduction-band offset of 2.25~eV. In a sandwich geometry, the polar discontinuity can be neglected through a proper choice of the terminations and the LaAlO$_3$/SrTi$_{1-x}$Nb$_x$O$_3$/LaAlO$_3$ heterostructure should offer a superconducting layer confined in a deep well of order 2~eV. Setting $U$ to 2.25~eV, we obtain the dashed line in Fig.~\ref{fig:fit}(b) with stronger quantum oscillations. If it turns out to be feasible, this system would allow one to explore lower values of $x$, for which the relative $T_c$ boost is theoretically even larger [see Fig.~\ref{fig:map}(a)].

\section{Conclusion}

We have grown epitaxial thin films of superconducting Nb-doped STO embedded in pristine insulating STO and measured their resistance and Hall carrier density with the prospect of seeing shape resonances, i.e, quantum oscillations and increase of $T_c$ with reducing film thickness. Except for the thinnest films that are insulating, our measurements confirm the expectations based on a model of confined BCS superconductor, where the confinement stems from the screened potential of Nb dopants. Semi-quantitative agreement between theory and experiment is achieved by averaging the calculated $T_c$ over a distribution of carrier densities whose width equals the measured fluctuations of the Hall density among different samples. This highlights the possible presence of density inhomogeneity in the films. Our results provide the first analysis of $T_c$ modulation by quantum confinement in epitaxial superconducting thin films at very low carrier density. Based on the model, we propose to enhance the quantum size effects further by either increasing Nb doping, which makes the confinement deeper, or by changing substrate and capping layers in view of a harder confinement, and then reducing the Nb doping to reach even lower densities where the shape resonances are amplified.

\section{Methods}

\subsection{Sample preparation}
SrTi$_{1-x}$Nb$_x$O$_3$ (Nb:STO) films at 1 atomic \% Nb with various thicknesses were grown on $5\times 5$~mm$^2$ TiO$_2$-terminated (001)-oriented SrTiO$_3$ (STO) substrates (Crystec GmbH) by pulsed laser deposition at $1100\,^\circ$C in an oxygen pressure of $10^{-6}$~Torr. A $40$~u.c.\ undoped STO capping layer was then grown on top of the Nb:STO layer at $800\,^\circ$C in an oxygen pressure of $8\times10^{-5}$~Torr, to form a symmetric confinement potential for the doped layer. During the deposition, the laser fluence was set to approximately $0.6$~J~cm$^{-2}$ and the repetition rate was kept at $1$~Hz. The deposition was fully monitored by RHEED and specular spot intensity oscillations. The growth rate was approximately $50$ laser pulses per Nb:STO monolayer. After growth, all the samples were annealed at $550\,^\circ$C in $200$~mbar of O$_2$ for one hour, before the final cooling to room temperature in the same atmosphere.

\subsection{Transport measurement}
Electric transport was measured using a standard four-probe method with a van der Pauw configuration. Aluminum wires were directly ultra-sonically bonded to the sample to form ohmic contacts. Superconductivity measurements were performed in a $^3$He/$^4$He dilution refrigerator. Magnetic fields up to $7$~T were applied perpendicular to the samples to measure the Hall resistance. Table~\ref{table} reports the measured sheet resistance $R_s$ at 600~mK (except for 19~u.c., where the temperature was 496~mK), as well as the measured Hall density and the resulting elastic scattering time $\tau=m/(n_{\mathrm{H}}e^2R_s)$ for a mass of one electron mass.

\begin{table}[t]
\caption{\label{table}
Sheet resistance $R_s$, Hall density $n_{\mathrm{H}}$, and elastic scattering time $\tau$ for various thicknesses $L$.
}
\begin{tabular*}{\columnwidth}{c@{\extracolsep{\fill}}ccc@{\extracolsep{0em}}c}
\hline\hline
$L$ (u.c.) & $R_s$ ($\Omega$) & $n_{\mathrm{H}}$ (10$^{13}$~cm$^{-2}$) & $\tau$ (10$^{-12}$~s) \\
\hline
11 & 41.87 & 4.22 & 2.0 \\
13 & 19.73 & ---  &     \\
14 & 57.78 & 4.49 & 1.4 \\
15 & 14.29 & ---  &     \\
17 & 31.70 & 6.89 & 1.6 \\
19 & 36.60 & ---  &     \\
22 & 28.68 & 8.89 & 1.4 \\
\hline\hline
\end{tabular*}
\end{table}

\section{Acknowledgements}
This work was supported by the Swiss National Science Foundation (SNSF) through project 200021 – 162628 and through the SNSF Early Postdoc Mobility Grant P2GEP2$\_$181450 (D.\@ V.\@). The authors also acknowledge the support of the Swiss National Science Foundation through Division II and of the European Research Council under the European Union’s Seventh Framework Program (FP7/2007-2013)/ERC Grant Agreement No.\@ 319286 (Q-MAC).

\end{document}